\documentclass[prl,10pt,aps,twocolumn,showpacs,floatfix,amssymb,amsmath]{revtex4-1}
\usepackage{slashed}
\usepackage{epsfig}

\newcommand{\beqn}{\begin{eqnarray}}
\newcommand{\eeqn}{\end{eqnarray}}
\newcommand{\eq}[1]{(\ref{#1})}

\newcommand{\Z}{{Z \!\!\! Z}}

\begin{document}

\title{Comment on ``Charged vector mesons in a strong magnetic field''}

\author{M.~N.~Chernodub}\thanks{On leave from ITEP, Moscow, Russia.}
\affiliation{CNRS, Laboratoire de Math\'ematiques et Physique Th\'eorique, Universit\'e Fran\c{c}ois-Rabelais Tours,\\ F\'ed\'eration Denis Poisson, Parc de Grandmont, 37200 Tours, France}
\affiliation{Department of Physics and Astronomy, University of Gent, Krijgslaan 281, S9, B-9000 Gent, Belgium}

\begin{abstract}
In a recent paper Y.~Hidaka and A.~Yamamoto [Phys.\ Rev.\ D  87 (2013) 094502] claim -- using both analytical and numerical approaches -- that the charged $\rho$ mesons cannot condense in the vacuum subjected to a strong magnetic field. In this Comment we point out that both analytical and numerical results of this paper are consistent with the {\it inhomogeneous} $\rho$-meson condensation. Furthermore, we show that the numerical results of the paper support the presence of the expected (in quenched lattice QCD) crossover transition driven by the $\rho$--meson condensation. Finally, we stress that the inhomogeneous $\rho$--meson condensation is consistent with both Vafa-Witten and Elitzur theorems.
\end{abstract}



\date{September 16, 2013}

\maketitle

\noindent
{\it Vafa-Witten theorem and $\rho$--meson condensation.} \\[1mm]
The authors of Ref.~\cite{Hidaka:2012mz} present analytical and numerical arguments demonstrating that the charged $\rho$ mesons cannot condense spontaneously in the vacuum of QCD subjected to a strong magnetic field background. 

The analytical part of Ref.~\cite{Hidaka:2012mz} claims that the vector meson condensation cannot occur because the presence of these condensates would break the diagonal subgroup of the global isospin group $U(1)_{I_3}$ of QCD and lead to the appearance of a massless Nambu-Goldstone boson in the spectrum of the theory in contradiction with the Vafa-Witten theorem~\cite{Vafa:1983tf}. 

However, the internal symmetries of the discussed system correspond to a larger theory, QCD$\times$QED because QCD in the background of a strong magnetic field is evidently coupled to electromagnetism. The global isospin group $U(1)_{I_3}$ of QCD is a part of the $U(1)_{\mathrm{em}}$ gauge group of QED, so that the would-be Nambu-Goldstone boson should inevitably be absorbed by (a component of) the electromagnetic field via an analogue of the Higgs mechanism. Consequently, no massless particles -- that would signal the breaking of the global $U(1)_{I_3}$ symmetry -- should appear in the spectrum~\cite{Chernodub:2012zx} (a related discussion in a toy model may also be found in Ref.~\cite{Li:2013aa}).

Notice that in most approaches to QCD in a magnetic field background, the latter is introduced as a background classical field in a fixed electromagnetic $U(1)_{\mathrm{em}}$ gauge. The would-be Nambu-Goldstone modes should disappear from the spectrum due to a Higgs mechanism which, as a physical phenomenon, works regardless of the fact whether the gauge is fixed or not, both in fixed (classical) and dynamical backgrounds. Thus, the condensation of the $\rho$ mesons in a fixed magnetic background should not lead to appearance of a Nambu-Goldstone boson in agreement with the Vafa-Witten theorem~\cite{Vafa:1983tf}.

\vskip 3mm
\noindent
{\it Inhomogeneities of the $\rho$--meson condensate.}\\[1mm]
The numerical part of Ref.~\cite{Hidaka:2012mz} supports the analytical part of the same paper by demonstrating that the two-point correlation functions of the $\rho$ meson fields vanish at large separations in quenched lattice QCD simulations:
\beqn
\lim_{|x - y| \to \infty} \langle \rho^\dagger(x) \rho(y) \rangle_{\mathrm{lattice}} = 0\,.
\label{eq:correlation:lat}
\eeqn
In this Comment we show that the numerical results of Ref.~\cite{Hidaka:2012mz} are consistent with the inhomogeneous $\rho$-meson condensation because the asymptotic large-volume behavior of the two-point function~\eq{eq:correlation:lat} cannot be used to reveal the presence or absence of the $\rho$--meson condensation due to strong inherent inhomogeneity of the condensate predicted in Ref.~\cite{Chernodub:2010qx}.

The $\rho$--meson condensation in zero-temperature and zero-density QCD is an interesting phenomenon because it may correspond to a new phase of QCD characterized by a perfect electric conductivity (``superconductivity'') along the magnetic field axis~\cite{Chernodub:2010qx,Chernodub:2011mc}. In other words, the sufficiently strong magnetic field may turn the vacuum into a superconductor with zero electrical resistance~\cite{Chernodub:2012tf}.

In the mean--field approach the critical magnetic field of the vacuum insulator--superconductor transition is
\beqn
e B_c = m_\rho^2\,,
\label{eq:eBc}
\eeqn
where $m_\rho \equiv m_\rho(B=0)$ is the mass of the $\rho$ meson in the absence of the magnetic field. Quantum corrections rise the critical field~\eq{eq:eBc} to a higher value~\cite{Callebaut:2011ab}.

The electric superconductivity of the vacuum is caused by a $(p + i p)$--wave condensation of the charged $\rho$--meson field $\rho_\mu \equiv \langle {\bar u} \gamma_\mu d\rangle$ with $\rho = \rho_1 = - i \rho_2 \neq 0$. Other vector components of the condensate are zero: $\rho_0 = \rho_3 = 0$.

In the mean--field approach the condensate solution takes the following form~\cite{Chernodub:2011gs}:
\beqn
\rho_{\mathrm{MF}}(x_1, x_2) & = & \sum_{n \in \Z} C_n h_n \left(\nu, \frac{x_1 + i x_2}{L_B}, \frac{x_1 - i x_2}{L_B} \right),
\label{eq:rho:sol} \\
h_n(\nu, z, {\bar z}) & = & e^{ - \frac{\pi}{2} (|z|^2 + {\bar z}^2 ) - \pi \nu^2 n^2 + 2 \pi \nu n {\bar z}}\,,
\label{eq:h:sol}
\eeqn
where the choice of the constants $C_n$ and $\nu$,
\beqn
C_{n+2} = C_n\,, \qquad C_1 = i C_0\,, \qquad \nu = \frac{\sqrt[4]{3}}{\sqrt{2}}\,,
\label{eq:C:sol}
\eeqn
corresponds to a hexagonal (equilateral triangular) lattice pattern in the transverse $(x_1,x_2)$ plane,~and 
\beqn
L_B = \sqrt{\frac{2 \pi }{|e B|}}
\label{eq:L:B}
\eeqn
is the magnetic length. The magnetic field is directed along the third axis, $x_3$.

The parameter $C_0$ in Eq.~\eq{eq:C:sol} is fixed by the requirement of energy minimization: the $\rho$--meson condensate is zero in the hadronic phase at $B < B_c$ while at $B > B_c$ the ground state develops a nonzero condensate $\rho \sim C_0 \sim \sqrt{B - B_c}$ in order to lower the energy density of the ground state (details can be found in Ref.~\cite{Chernodub:2011gs}). 

The condensate~\eq{eq:rho:sol}, \eq{eq:h:sol}, \eq{eq:C:sol} is an anisotropic and inhomogeneous structure, which consists of an infinite number of the $\rho$ vortices parallel to the magnetic field. There is one $\rho$ vortex per unit area $L_B^2 \equiv 2 \pi / |e B|$ of the transverse plane. The analyses of the $\rho$-vortex lattice structure in an effective field model and in a holographic approach are given in Ref.~\cite{Chernodub:2011gs} and \cite{Bu:2012mq}, respectively. 

\begin{figure}[!thb]
\begin{center}
\includegraphics[scale=0.3,clip=false]{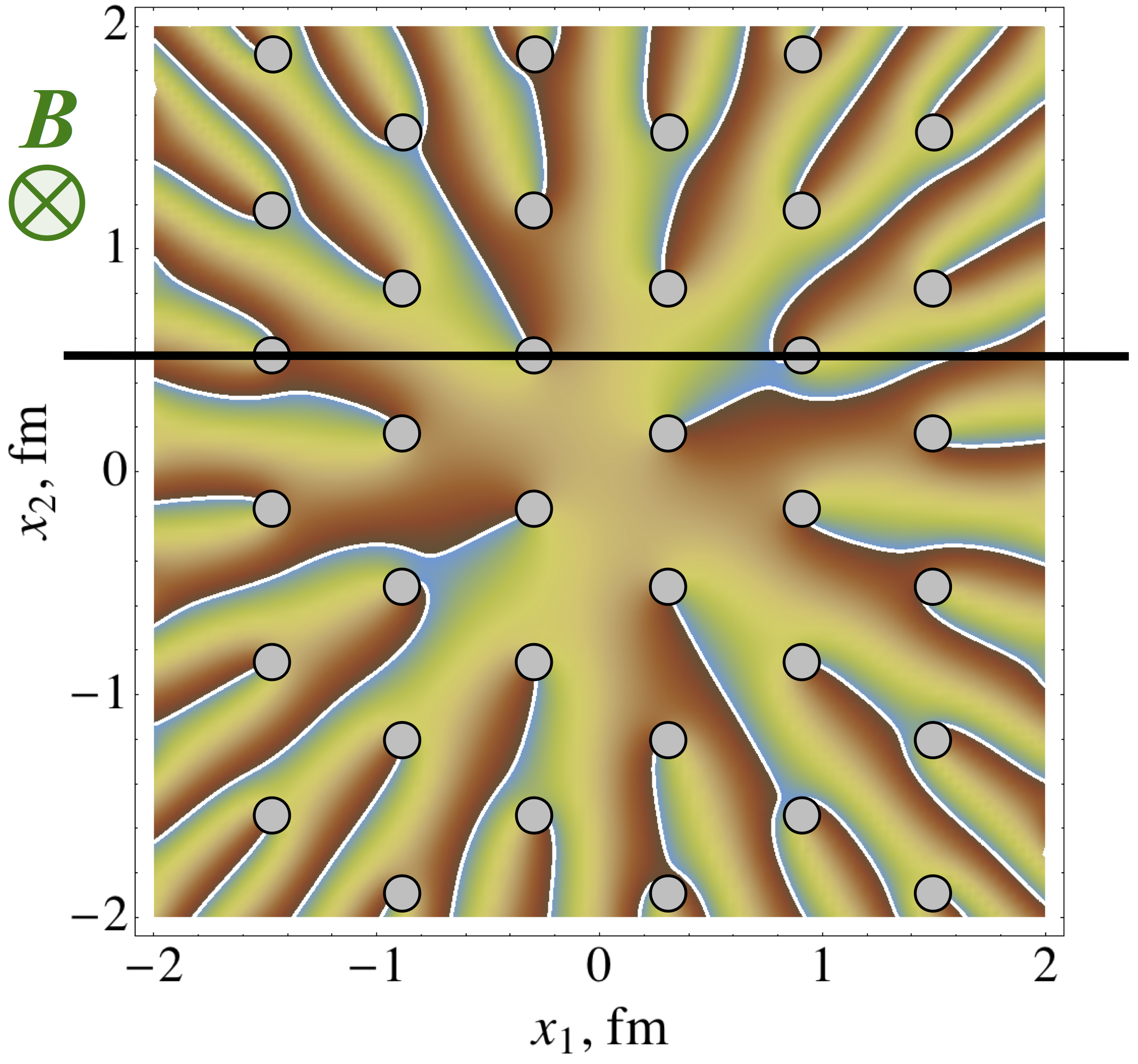} 
\end{center}
\vskip -5mm
\caption{The density plot of the phase $\varphi = \arg \rho_{\mathrm{MF}}(x_1,x_2)$ of the condensate solution~\eq{eq:rho:sol}, \eq{eq:h:sol}, \eq{eq:C:sol} in the $(x_1,x_2)$ plane perpendicular to the magnetic field $\boldsymbol B$ at $B = 1.01 B_c$. The red (darker) color corresponds to small phase ($\varphi \sim 0$) and blue (lighter) color to large values of the phase ($\varphi \sim 2\pi$). The white lines mark positions of the $2\pi$ cuts in the phase (Dirac sheets), while their endpoints, denoted by the gray circles, corresponds to the positions of the $\rho$ vortices. The  $\rho$ vortices are arranged in a hexagonal lattice~\cite{Chernodub:2011gs,Bu:2012mq}. The horizontal black line denotes the position of $(x_1,x_3)$ plane used in Fig.~\ref{fig:correlations}.}
\label{fig:phases}
\end{figure}

Similarly to the Abrikosov lattice states in type-II superconductors~\cite{ref:Abrikosov}, the $\rho$-meson condensate vanishes at the positions of the $\rho$ vortices while the phase of the $\rho$ field winds around the vortex cores. In particular, the inhomogeneity is reflected in the fact that the phase of the condensate is a rapidly oscillating function of the transverse coordinates, Fig.~\ref{fig:phases}. The latter fact is crucial for our analysis below.

~\\
\vskip 3mm
\noindent
{\it Vanishing zero-momentum component of the condensate.}\\[1mm]
Due to the inhomogeneity of the condensate (and, especially, of its phase), the bulk average of the $\rho$ condensate~\eq{eq:rho:sol}, \eq{eq:h:sol}, \eq{eq:C:sol} over the whole transverse $x_\perp \equiv (x_1,x_2)$ plane (and over whole space) is {\it always zero},
\beqn
\langle \rho \rangle 
\equiv \left\langle \frac{1}{{\mathrm{Vol}}_\perp} \int\! d^2 x_\perp \rho(x)\right\rangle 
\equiv \left\langle \frac{1}{{\mathrm{Vol}}} \int\! d^4 x \, \rho(x)\right\rangle 
\equiv 0. \quad
\label{eq:rho:0}
\eeqn
In other words, the ${\boldsymbol p}_\perp \equiv (p_1,p_2) = 0$ component of the $\rho$-meson condensate should always be vanishing in the ground state, if even the condensate itself is nonzero. 

Due to the translational symmetry of QCD, Eq.~\eq{eq:rho:0} implies that the expectation value of the local operator $\rho(x)$ should also be vanishing in a finite physical volume. Indeed, all coordinate-shifted copies of any field configuration enter the partition function with the same weight so that the vacuum expectation value (v.e.v.) of the local field operator $\left\langle \rho(x) \right\rangle$ is equal to the v.e.v. of its average over the whole space. The latter is zero~\eq{eq:rho:0} in agreement with the Elitzur's theorem~\cite{ref:Elitzur}. Thus, on a practical side, $\langle \rho(x) \rangle$ is not a good local order of the inhomogeneous $\rho$-meson condensation. Notice that the homogeneous condensate of $\rho$ mesons in QCD is ruled out both by Ref.~\cite{Hidaka:2012mz} and Ref.~\cite{Chernodub:2011mc}.

It is worth noticing that very same property~\eq{eq:rho:0} is shared by the celebrated Abrikosov vortex lattices in type-II superconductors~\cite{ref:Abrikosov}: despite the vortex--lattice ground state is a superconducting state with a locally large order parameter~\eq{eq:rho:sol}, the bulk average of the corresponding order parameter is nevertheless vanishing due to the unavoidable presence of the Abrikosov vortices.

\begin{figure}[!thb]
\begin{center}
\includegraphics[scale=0.45,clip=false]{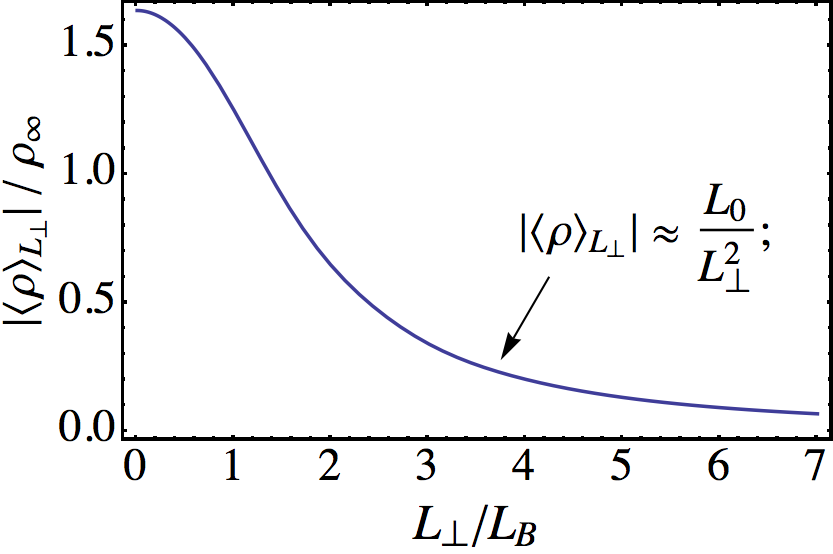} 
\end{center}
\vskip -5mm
\caption{The $\rho$ condensate~\eq{eq:rho:sol}, \eq{eq:h:sol}, \eq{eq:C:sol} averaged over the $L_\perp^2$ area in the transverse plane~\eq{eq:rho:L} as a function of the magnetic length~$L_B$, Eq.~\eq{eq:L:B}. The quantity $\rho_\infty$ is given in Eq.~\eq{eq:rho:inf} and the $L_\perp \to \infty$ asymptotic is calculated in Eq.~\eq{eq:rho:volume}.}
\label{fig:rho:L}
\end{figure}

In order to make our statements more quantitative, let us consider the mean value of the condensate~\eq{eq:rho:sol}, \eq{eq:h:sol}, \eq{eq:C:sol} in a $L_\perp \times L_\perp$ area of the transverse plane:
\beqn
\langle \rho \rangle_{L_\perp} = \frac{1}{L^2_\perp} 
\iint\limits_{- L_\perp/2}^{\quad\ L_\perp/2} \!\!\! d x_1 d x_2
\ \rho(x_1,x_2)\,. \quad
\label{eq:rho:L}
\eeqn
This integral can be calculated numerically, as shown in Fig.~\ref{fig:rho:L}. The condensate~\eq{eq:rho:sol}, \eq{eq:h:sol}, \eq{eq:C:sol} is proportional to the prefactor $C_0$ which can also be related to the bulk average of the squared absolute value of the condensate:
\beqn
\rho_\infty^2 \equiv \lim_{L_\perp \to \infty} \langle |\rho|^2 \rangle_{L_\perp} = \frac{|C_0|^2}{2 \sqrt[4]{3}}\,.
\label{eq:rho:inf}
\eeqn

According to Fig.~\ref{fig:rho:L} the mean value of the inhomogeneous condensate vanishes quickly~\cite{footnote1} with the increase of the transverse area $L_\perp^2$. In the large-volume limit $L_\perp \gg L_B$ the average of the inhomogeneous condensate has the following asymptotic behavior:
\beqn
|\langle \rho \rangle_{L_\perp} | = \frac{L_0}{L^2_\perp}  + O\left(L^{-4}_\perp \right)\,,
\label{eq:rho:volume}
\eeqn
where $L_0 = \alpha L_B \rho_\infty$ and $\alpha \approx 3.27$ is a numerical constant associated with the hexagonal geometry of the $\rho$--vortex lattice.

\vskip 3mm
\noindent
{\it Correlation functions.}\\[1mm]
We have just seen that in the large volume the expectation value of the $\rho$-meson field cannot serve as a good order parameter of the {\it inhomogeneous} $\rho$-meson condensation. 
The same statement is true for the two--point correlation functions of the $\rho$--field operators in the condensed ground state: At a large separation the decoupling occurs and the result vanishes,
\beqn
\lim_{|x - y| \to \infty} \langle \rho^\dagger(x) \rho(y) \rangle_{\mathrm{theory}} =  |\langle \rho \rangle|^2 \equiv 0\,,
\label{eq:correlation}
\eeqn
in agreement with lattice results in a large volume~\eq{eq:correlation:lat}.

If the suspected condensate were homogeneous (i.e., independent of space-time coordinates), then Eq.~\eq{eq:correlation} would signal the absence of this condensate. However, as we have just seen, for the inhomogeneous condensate this statement is no more valid. In other words, the vanishing of the asymptotic correlation function~\eq{eq:correlation} is fully consistent with the inhomogeneous $\rho$--meson condensation.

\begin{figure}[!thb]
\begin{center}
\includegraphics[scale=0.35,clip=false]{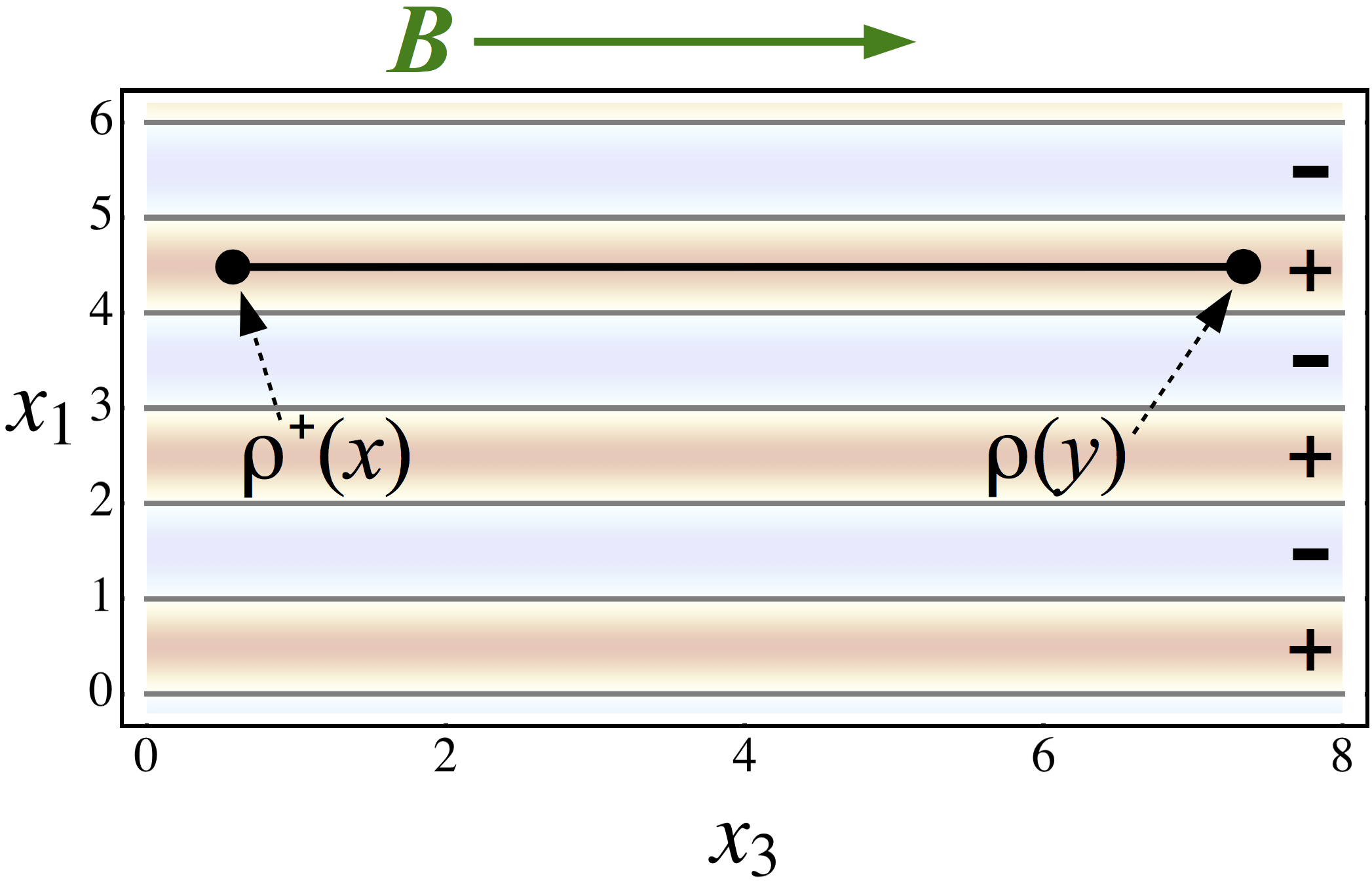} \\
\includegraphics[scale=0.35,clip=false]{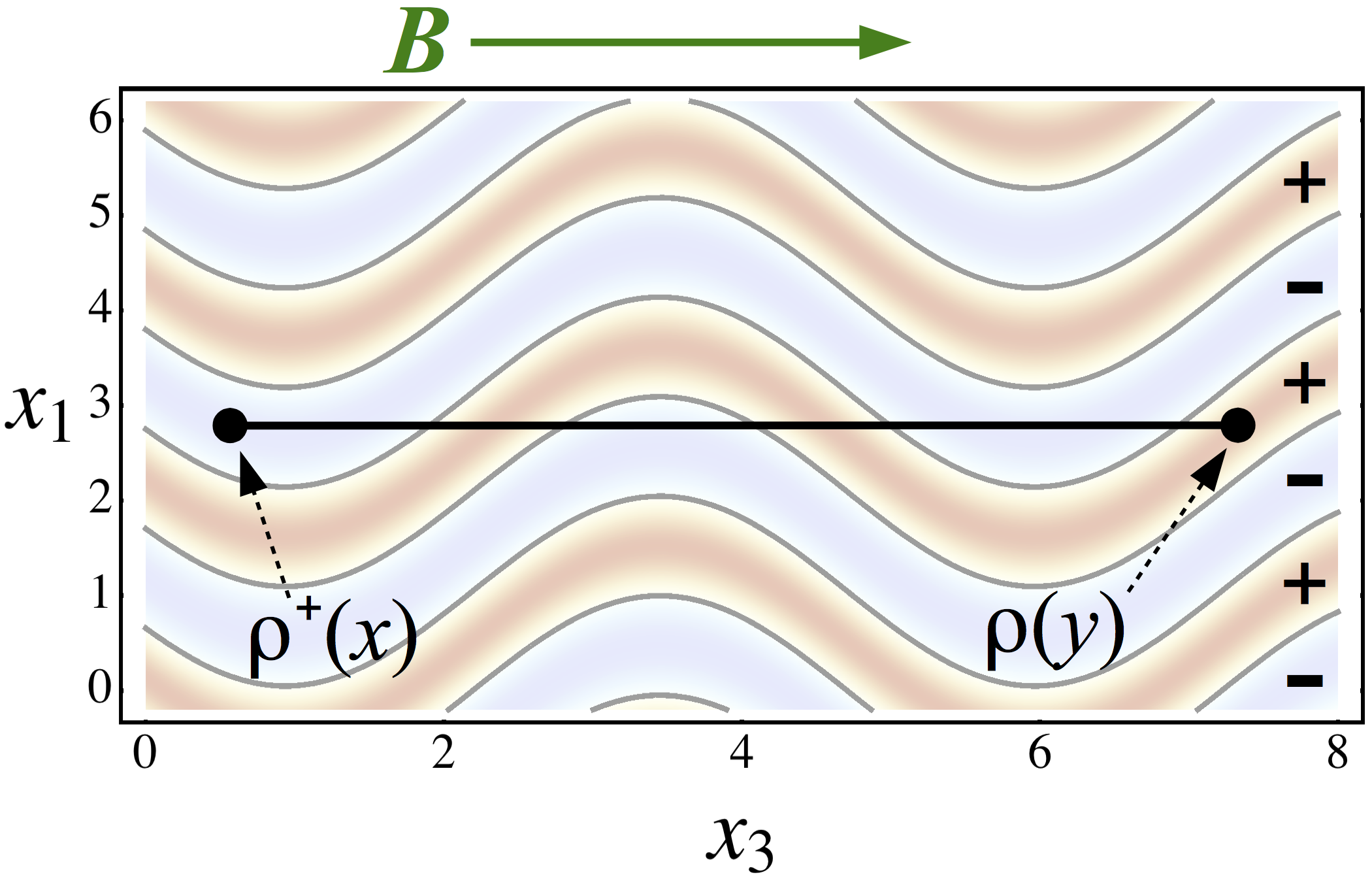} 
\end{center}
\vskip -5mm
\caption{Schematic illustration of the inhomogeneous $\rho$--meson condensate in the $(x_1,x_3)$ plane: (top) mean-field solution and (bottom) a real state with quantum fluctuations taken into account. The plane corresponds to the crosssection shown in Fig.~\ref{fig:phases} by a straight line. The red (darker) and blue (lighter) regions correspond to (predominantly) positive and negative values of the $\rho$ condensate, respectively. The gray lines show the positions of the vortices with $\rho = 0$ (given by the gray circles in Fig.~\ref{fig:phases}). The phase of $\rho$ is coherent [$\rho^\dagger(x_3) \rho (y_3) \equiv |\rho(x_3)|^2$] for the mean field solution while in the real vacuum this is no more the case, $\rho^\dagger(x_3) \rho (y_3) \neq |\rho|^2$. Notice that $\langle \rho(y) \rangle = 0$ due to the vortex fluctuations even in the presence of the inhomogeneous condensate. }
\label{fig:correlations}
\end{figure}

One could alternatively suggest that the condensate may be calculated in a special case of a two-point function with longitudinally separated points $x_3 \neq y_3$ (while $x_\mu = y_\mu$ for $\mu \neq 3$). Indeed, in this case the mean--field solution for the condensate~\eq{eq:rho:sol}, \eq{eq:h:sol} is independent of the longitudinal coordinate $x_3$ and, consequently, the $\rho$ vortices are parallel to the magnetic field, Fig.~\ref{fig:correlations}(top). Thus, for the mean-field solution the phases of the $\rho^\dagger$ and $\rho$ operators in the correlator $\langle \rho^\dagger (x)\rho (y)\rangle$ cancel each other and one could expect that the correlation function becomes equal to generally nonvanishing quantity $|\rho_{\mathrm{MF}}(x_1,x_2)|^2$ for asymptotically separated points $|x_3 - y_3 | \to \infty$. 

However, the mentioned mean-field arguments do not work in the real quantum system because the vortices are no more strictly parallel to the magnetic field axis due to inevitable vortex vibrations, Fig.~\ref{fig:correlations}(bottom). The vortex vibrations were indeed observed in quenched lattice QCD in Ref.~\cite{Braguta:2013uc}. Thus, in the real system in a large volume, the coherent ``mean--field--like'' cancellation of the phases of the $\rho$--meson fields does not work and the decoupling of the expectation values~\eq{eq:correlation} holds true.

\vskip 3mm
\noindent
{\it Coherence length and condensate in a moderate volume.}\\[1mm]
One may expect the existence of certain decoherence length $l_c = l_c(B)$ related to the intrinsic rigidity of the vortices at given strength of the magnetic field. The length $l_c$ is defined as a maximal  length of a vortex segment at which the mean mutual fluctuations of positions of the segment's ends in transverse directions are equal to the average inter-vortex distance, $\langle (x_\perp - y_\perp)^2 \rangle \sim L_B^2$. The vortex segments which are shorter than $l_c$ are basically straight lines which are approximately parallel to the magnetic field axis. Thus, at the moderate length scales $|x_3 - y_3| \lesssim l_c$ the quantum vortex state resembles the straight mean-field result~\eq{eq:rho:sol}.

As a consequence, in moderate space volumes $L^3$ with $L \sim l_c$ the vortex state may be quite close to the mean-field solution, Fig.~\ref{fig:correlations}(top). Thus, the correlation function along the magnetic field should give us a nonzero mean-field result at the maximal available point separation, $|x_3 - y_3| = L/2$:  $\langle\rho^\dagger(x_\perp,0) \rho (x_\perp,L/2) \approx |\rho_{\mathrm{MF}}(x_\perp,0)|^2$. 

A nonzero $\rho$--meson condensate was indeed found numerically in simulations of quenched lattice QCD at moderate volumes~\cite{Braguta:2011hq}. At larger volumes (studied in Ref.~\cite{Hidaka:2012mz}) the decoupling~\eq{eq:correlation} should happen and the two-point correlations functions cannot be used to reveal the presence of absence of the inhomogeneous condensation.

\begin{figure}[!thb]
\begin{center}
\includegraphics[scale=0.45,clip=false]{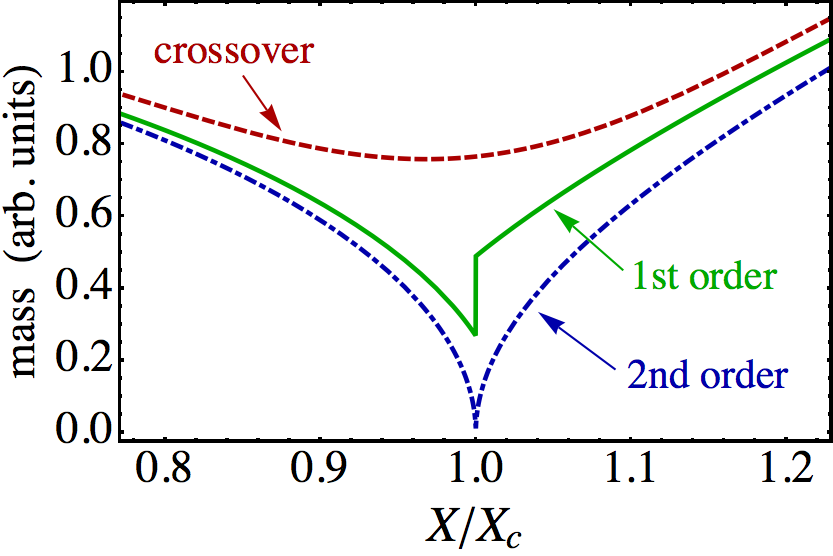} 
\end{center}
\vskip -5mm
\caption{Qualitative behavior of a mass of a lowest excitation associated with an order parameter in a generic system as a function of a thermodynamic parameter $X$ (magnetic field $B$, temperature $T$ etc) for a first and second order transitions and for a crossover. }
\label{fig:mass}
\end{figure}

\vskip 3mm
\noindent
{\it Mass of the $\rho$ meson excitation.}\\[1mm]
The mean-field calculations~\cite{Chernodub:2010qx,Chernodub:2011mc,Chernodub:2011gs,Callebaut:2011ab} predict that the transition to the superconducting phase should be of the second order: the mass of the lowest $\rho$-meson excitation should vanish at the critical field~\eq{eq:eBc}. One may expect that the inclusion of the quantum fluctuations may enhance or weaken the transition making it either a first order transition or a crossover, respectively. Notice that in the cases of the first-order transition and crossover the $\rho$--meson mass should not be vanishing at the transition point. An illustration of a generic behavior of the lowest mass for all these transitions is shown in Fig.~\ref{fig:mass}.

For example, both first, second and crossover transitions are realized in the electroweak model at a finite temperature. The strength of the transition depends on the value of the zero--temperature Higgs mass. In this model the behavior of the lowest (scalar) mass on temperature $T$ follows Fig.~\ref{fig:mass} (with $X \equiv T$)~\cite{ref:EW}.

Contrary to the real QCD with dynamical light fermions, in the zero-temperature quenched lattice QCD the magnetic-field-induced transition should always be of the crossover type. Indeed, the gluons do not couple to the electromagnetic field directly while the dynamical fermions are absent in the quenched QCD, so that variation of the external magnetic field cannot lead to thermodynamic singularities in this theory. Given also the experience with the crossover transition in the electroweak model~\cite{ref:EW}, we expect that the dependence of the $\rho$ meson mass on the magnetic field in the quenched QCD should be similar to the one illustrated by the dashed line in Fig.~\ref{fig:mass}. Not surprisingly, the quenched result (Fig.~1 of Ref.~\cite{Hidaka:2012mz}) on the $\rho$-meson mass confirms our expectation.

\vskip 5mm
\noindent
{\it Conclusions.}\\[1mm]
In Ref.~\cite{Hidaka:2012mz} it was claimed that the charged $\rho$ mesons cannot condense in strong magnetic field.

We point out that the analytical and numerical results of Ref.~\cite{Hidaka:2012mz} are consistent with the inhomogeneous $\rho$-meson condensation predicted in Refs.~\cite{Chernodub:2010qx,Chernodub:2011mc}. In particular, a large-volume limit of the two-point correlation function calculated in Ref.~\cite{Hidaka:2012mz} cannot be used to support the absence of the inhomogeneous $\rho$--meson condensation.

Moreover, we show that the results of Ref.~\cite{Hidaka:2012mz}~on 
\begin{itemize}
\item[(i)] the behavior of the asymptotic value of the correlation function as the function of the system volume, 
\item[(ii)] the behavior of the $\rho$--meson mass as the function of the magnetic field 
\end{itemize}
are, in fact, consistent with the expected crossover transition associated with the inhomogeneous $\rho$--meson condensation in quenched lattice QCD.

\vskip 3mm

The work is supported by Grant No. ANR-10-JCJC-0408 HYPERMAG (France). The author is grateful to Y.~Hidaka for extensive discussions.

\end{document}